\documentclass[aps,pra,superscriptaddress,amsmath,amssymb,showpacs,a4paper,twocolumn,10pt]{revtex4-1}
\usepackage{dsfont, color, graphics,yfonts}
\usepackage{braket}
\usepackage{amsmath,amssymb}

\usepackage{graphicx,epsfig}

\input epsf

\def\beq{\begin{equation}}
\def\eeq{\end{equation}}
\def\beqa{\begin{eqnarray}}
\def\eeqa{\end{eqnarray}}
\def\bra#1{\mathinner{\langle{#1}|}}
\def\ket#1{\mathinner{|{#1}\rangle}}
\def\mn#1{\langle #1 \rangle}
\def\prjct#1{\mathinner{|{#1}\rangle}\!\!\mathinner{\langle{#1}|}}

\def\eq{\begin{equation}}
\def\eeq{\end{equation}}
\def\beqa{\begin{eqnarray}}
\def\eeqa{\end{eqnarray}}
\def\prjct#1{\mathinner{|{#1}\rangle}\!\!\mathinner{\langle{#1}|}}

\newcommand{\fig}[1]{Fig.~#1}
\def\tr{\text{Tr}}

\newcommand{\coh}[2]{\mathinner{|{#1}\rangle}\!\!\mathinner{\langle{#2}|}}

\begin{document}

\title{Testing unconventional decoherence models with atoms in optical lattices}

\author{Ji\v{r}\'{i} Min\'{a}\v{r}}
\affiliation{School of Physics and Astronomy, University of Nottingham, Nottingham NG7 2RD, United Kingdom}
\author{Pavel Sekatski}
\affiliation{Institut for Theoretische Physik, Universitat of Innsbruck, Technikerstra{\ss}e 25, A-6020 Innsbruck, Austria}
\author{Robin Stevenson}
\affiliation{School of Physics and Astronomy, University of Nottingham, Nottingham NG7 2RD, United Kingdom}
\author{Nicolas Sangouard}
\affiliation{Department of Physics, University of Basel, Klingelbergstrasse 82, 4056 Basel, Switzerland}

\pacs{
03.65.Yz, 
42.50.Xa, 
37.10.Jk, 
67.85.d 
}

\begin{abstract}
  Various models have been proposed in which the Schr\"{o}dinger equation is modified to account for a decay of spatial coherences of massive objects. While optomechanical systems and matter-wave interferometry with large clusters are promising candidates to test these models, we here show that using available techniques for atoms in optical lattices, some of these models can be efficiently tested. In particular, we compare unconventional decoherence due to quantum gravity as introduced by Ellis and co-workers [Phys. Lett. B {\bf 221}, 113 (1989)] and conventional decoherence due to scattering of the lattice photons and conclude that optimal performances are achieved with a few atoms in the realistic case where product atomic states are prepared. A detailed analysis shows that a single atom delocalized on a scale of 10 cm for about a second can be used to test efficiently the hypothetical quantum gravity induced decoherence.
\end{abstract}

\maketitle


\noindent \emph{Introduction --} By virtue of the linearity of quantum theory, a system like an atom can be in a superposition of two different positions, as shown e.g. in atom interferometry \cite{Gould_1986, Keith_1988, Borde_1989}. This superposition principle is supposed to hold for more massive systems as quantum theory makes no distinction between small and large systems. To better account for what we observe at macroscopic scales, post-quantum models have been proposed in which quantum theory is supplemented with explicit collapse mechanisms \cite{Bassi_2013}. Various systems are being investigated to test these hypothetical models. Impressive experiments are being performed, for example in matter-wave interferometry with free falling atoms in which e.g. the wave packet of a single Rb atom gets separated by 1.4 centimeters for 2.3 seconds \cite{Dickerson_2013}. Since the mass in these experiments is limited to that of a single atom, matter-wave interferometry with larger and larger molecules and clusters is being developed \cite{Arndt_1999, Haslinger_2013}. Ref. \cite{Eibenberger_2013} is an example of ongoing experiment using a molecule made with more than 800 atoms with a molecular weight exceeding a hundred times the one of a single Rubidium atom. Optomechanical systems where heavy nano and micro mechanical oscillators are driven through the radiation pressure or dielectric nanoparticles levitating in the focus of intense laser fields are also at the core of intense research programs \cite{Aspelmeyer_2014}. While all these techniques are promising to test the superposition principle with massive systems in incoming years, it is natural to ask whether well-controlled and easy-to-use systems could be used to test these wave-function collapse models. 

While the ready-to-use toolbox to manipulate individual atoms trapped in optical lattices makes them very attractive, it is usually suggested that the mass of a single atom is too small and the possibility to control millions of atoms does not overcome the problem as they cannot be prepared in GHZ-like states in practice (which would mimic the superposition of the center of mass of the heavier composite systems). Here we challenge these preconceptions by analyzing in detail the capability of cold atoms in optical lattices to test collapse models. The analysis is performed by taking into account the dominant source of decoherence, namely the photon scattering from the trap lasers. We compare this standard decoherence to the unconventional decoherence model first introduced in Ref. \cite{Ellis_1984} and further elaborated in Refs. \cite{Ellis_1989,Ellis_1992}, suggesting that the coupling to the topologically non trivial spacetime configurations admitted by the underlying theory of quantum gravity, termed as wormholes, leads to the decay of spatial coherences of a test massive system. We show that the effects of these decoherence mechanisms are qualitatively different: while the former operates locally, the latter induces a collective noise on the state of the atoms. Focusing on the experimentally relevant case where the atoms are prepared in product states, we find that optimal performances are achieved with small atom numbers. A detailed analysis shows that a single rubidium atom delocalized over $\sim 10$ cm can be used to test the hypothetical decoherence due to quantum gravity \cite{Ellis_1989}. While we take this unconventional decoherence model as an example, the proposed system might be used to probe other collapse models, such as the Diosi-Penrose model \cite{Diosi_1989, Penrose_1996}, the GRW model \cite{Ghirardi_1986, Ghirardi_1990, Gisin_1989} or a decoherence due to chameleon fields relevant for cosmology, where it is actually favorable to operate on small masses \cite{Burrage_2015}.\\

\noindent \emph{Principle --} We consider atoms trapped in a state dependent optical lattice, i.e. two ground states $|g\rangle$, $|s\rangle$ are trapped independently such that an atom prepared in a superposition $|g\rangle + |s\rangle$ can be spatially delocalized into $|g, x_g\rangle + |s, x_s\rangle,$ see Fig. 1a,b. To measure the decay of spatial coherences, various techniques can be envisioned depending e.g. on the atom number. In the case of a single atom, the atom is spatially relocalized and the coherence between the internal states is measured. To record the decay of spatial coherences in time, the atoms are kept in the state $|g, x_g\rangle + |s, x_s\rangle$ for longer times before relocalization and measurement of the $|g\rangle$ -- $|s\rangle$ coherence. Comparing the timescales of both the standard decoherence resulting form the scattering of the photons from the trap and unconventional decoherence models with the observed coherence decay time might make it possible to confirm or rule out predictions from unconventional decoherence mechanisms. \\

\begin{figure}[h!]	  
\includegraphics[width=8cm]{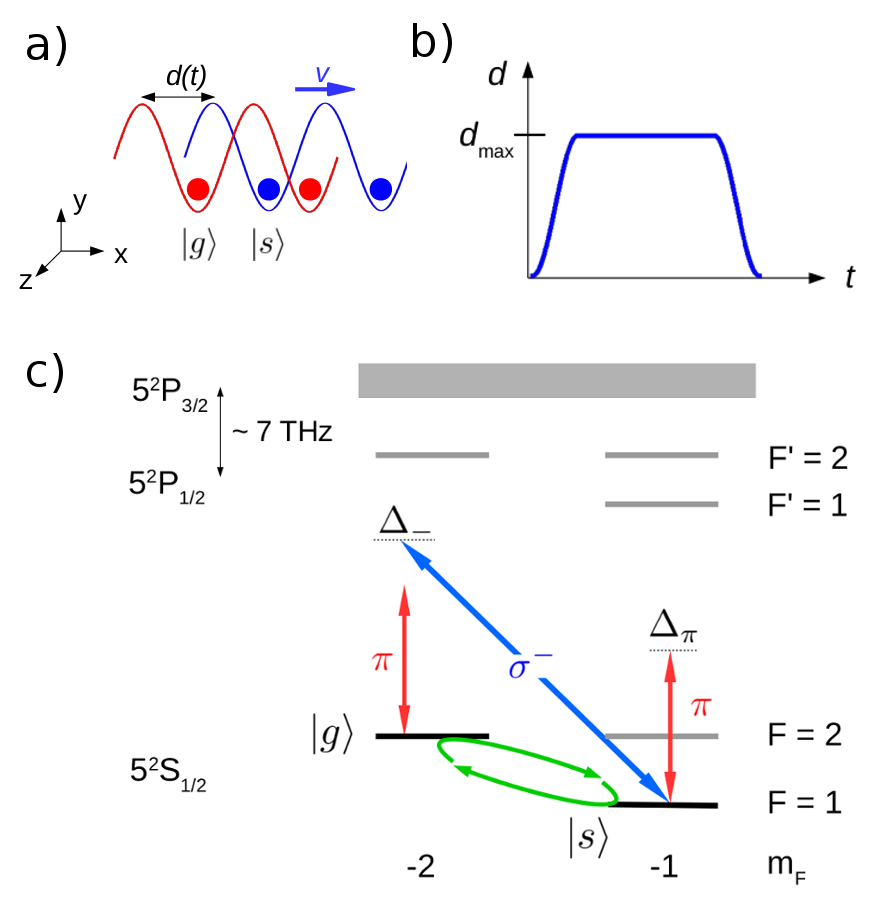}	  
\caption{(Color online) a) Schematic of the trapping potentials: Atoms in state $\ket{s}$ (blue circles) are separated by a distance $d(t)$ with respect to the atoms in state $\ket{g}$ (red circles). b) Time evolution of the separation $d$ between the $\ket{g}$ and $\ket{s}$ states. c) Level scheme of $^{87}$Rb. Atoms in $\ket{g}$ state are traped by the $\pi$ polarized trapping lasers (red arrows), while the $\ket{s}$ states are trapped by both the $\pi$ and $\sigma^-$ polarized (blue arrow) laser beams. Green arrows denote the microwave driving.} 
\label{fig:level scheme}  	
\end{figure}

\noindent \emph{State dependent manipulation of atomic motion --} We here show how to coherently manipulate the position of atoms in state-dependent atomic lattices. For concreteness, we focus on a far-off resonant dipole trapping of $^{87}$Rb \cite{Steck_2001, Sansonetti_2006}. The relevant level structure is shown in Fig. \ref{fig:level scheme}c. State dependent manipulation of atomic position can be obtained if $|g\rangle$ and $|s\rangle$ are chosen such that each state can be trapped with different polarizations. The transition $|g\rangle$ -- $|s\rangle$ also needs to be addressed coherently for internal state preparation and measurement. For example, one can identify $\ket{g} \equiv \ket{F=2,m_F=-2}$ and $\ket{s} \equiv \ket{F=1, m_F = -1}$. The detunings of the trapping lasers with respect to the P$_{1/2}$ manifold are $\Delta_-, \Delta_\pi$ for the $\sigma^-$ and $\pi$ polarizations respectively. We should note that the ground states couple to higher lying manifolds, namely the P$_{3/2}$ manifold. However it is easy to be in the regime, where the dominant contribution to the trapping potential comes only from the coupling to the P$_{1/2}$ manifold. For example if $\Delta_{-,\pi}$ are of order 100 GHz, the coupling to the P$_{3/2}$ manifold becomes negligible compared to the coupling to the P$_{1/2}$ manifold as P$_{3/2}$ and P$_{1/2}$ differ by 7 THz (see \fig{\ref{fig:level scheme}c}). In the following we work in a regime where $\Delta_- \approx \Delta_\pi \equiv \Delta$ and consequently for the lattice laser wavevectors $k_\pi \approx k_- = k$. Considering a one-dimensional geometry, where $x$ is the lattice axis, the trapping potentials for the two ground states are given by \cite[p. 199]{Foot_2005}

\begin{subequations}
\label{eq:Vlatt}
\begin{align}
	V_g(x,t) &= V^0_\pi \cos^2 \left( k x \right) \\
	V_s(x,t) &= V^0_\pi \cos^2 \left( k x \right) + V^0_- \cos^2 \left( k x + \varphi(t) \right), \label{eq:V_s}	
\end{align}
\end{subequations}
where $V^0_j = \hbar \Omega_j^2/(4\Delta)$, $j=\pi,-$, and $\varphi(t)$ is a time dependent offset of the $\sigma^-$ lattice which in practice can be achieved e.g. by varying the frequencies of the counter-propagating lattice beams \cite{Schmid_2006}. This leads to a time dependent separation $d(t)=\varphi(t)/k$ of the $\ket{g},\ket{s}$ states, see Fig. \ref{fig:level scheme}a. Note that we have absorbed possible multiplicities coming from the coupling of the $\ket{g},\ket{s}$ states to several levels into the Rabi frequencies.
\\

\noindent \emph{Decoherence --} In order to evaluate the decay of coherences in time, we compute the overlap of the initial and later time density matrices $O={\rm Tr}(\rho(0) \rho(t))$. While other figures of merit might be chosen, the overlap is a quantity that is easy to calculate and analytical expressions are given in the following sections. Moreover, in the case of a single atom, it yields directly the coherence decay rate. \\

\noindent \emph{Decoherence due to photon scattering: local dephasing --} While the internal state coherence can be degraded by many technical issues including lattice depth or magnetic field gradient fluctuations (see \cite{Alberti_2014, Schnorrberger_2009,Dudin_2013,Lisdat_2009} for an extensive study of decoherence mechanisms) we focus on the dominant (and unavoidable) source of decoherence, namely the scattering of lattice photons.

Let us consider first a single atom in a spatial superposition with separation $d$ between the superposed states. For short spatial separation $d \ll \lambda$ as compared to the wavelength of the trapping light, the rate of decoherence due to the scattering of the electromagnetic radiation grows quadratically with the distance 
\cite{Joos_1985}. 
For larger distances, which are relevant for testing unconventional decoherence, the decoherence rate saturates as predicted in \cite{Gallis_1990} and verified experimentally e.g. in \cite{Hackermuller_2004, Kokorowski_2001}. 
In particular, for lasers far detuned from resonance, it is given by 
\cite[p.180]{Foot_2005} 
\begin{subequations}
\label{eq:decay conv}
\begin{align}
	\Gamma_g &= \frac{1}{4} \frac{\Gamma_0}{2} \left( \frac{\Omega_\pi}{\Delta} \right)^2 \\
  \Gamma_s &= \frac{1}{4} \frac{\Gamma_0}{2} \frac{\Omega^2_\pi + \Omega_-^2}{\Delta^2},
\end{align}
\end{subequations}
for the $\ket{g},\ket{s}$ state respectively, $\Gamma_0$ being the atomic free space decay rate. Now, consider the case with N atoms where $d \gg \lambda$ and where the position of each atom is fully resolved through the scattered photons.  
The master equation governing the decoherence of $N$ atoms is given by a sum of independent scattering processes (we neglect the coherent evolution as we are interested solely in the decay of the coherences)
\beq
\dot \rho = \sum_{i=1}^N \sum_{\alpha \in\{g,s\}} \frac{\Gamma_\alpha}{2}\big(2 \prjct{\alpha}_i \rho \prjct{\alpha}_i -\{\prjct{\alpha}_i,\rho\} \big).
\eeq 
This equation can be rewritten as a \emph{local} dephasing process
\beq
  \dot{\rho} = \frac{\Gamma_{\rm sc}}{2} \left( \sum_{i=1}^N \sigma^i_z \rho \sigma^i_z - \rho \right),
  \label{eq:Master Eq local}
\eeq
where $\Gamma_{\rm sc}=\Gamma_g + \Gamma_s$ and $\sigma^i_z = \prjct{s}_k - \prjct{g}_k$ is the usual Pauli matrix in the $\{\ket{s},\ket{g}\}$ basis. Importantly, the decoherence due to photon scattering is independent of the separation $d(t)$. We can now compute the overlap of the initial and later time density matrices. Considering the initial atomic state to be a product state
\beq
	\ket{\psi_0}=\left(\frac{\ket{g}+\ket{s}}{\sqrt{2}}\right)^{\otimes N},
	\label{eq:prod state}
\eeq
it reads (see Appendix \ref{sec:Decoherence})
\beq
	O_\text{sc}(t)=\tr( \rho(0) \rho(t)) = \left(\frac{1+{\rm e}^{-(\Gamma_g + \Gamma_s)t } }{2} \right)^N.
	\label{eq:overlap conv}
\eeq  
\\

\noindent \emph{Unconventional decoherence: global dephasing --} Phenomenologically, the spatial decoherence can be described by
the master equation \cite{Bassi_2013}
\beq
	\dot \rho  =  - \int d{\bf x'} d{\bf x} \,\Gamma({\bf x}, {\bf x'}) \coh{\bf x}{\bf x} \rho \coh{\bf x'}{\bf x'}
	\label{eq:rho dot}
\eeq
where $\ket{{\bf x}}=\ket{x_1,...,x_N}$  is the position basis for the $N$ particles. The localization rate $\Gamma({\bf x}, {\bf x'})$ can be written as
\beq
 \Gamma({\bf x}, {\bf x'}) = \frac{\gamma_0}{2} \sum_{i,j=1}^N \mu_i \mu_j  \Big( \Phi(x_i - x_j)+\Phi(x_i' - x_j')- 2\Phi(x_i - x_j')\Big),
 \label{eq:Gamma}
\eeq
where the spatial dependence described by $\Phi$ (the so-called localization function) and the constants $\gamma_0$ and $\mu_i$ are given by the underlying microscopic theory. 

Following the treatment given in Refs. \cite{Ellis_1989, RomeroIsart_2011,Pepper_2012}, we now take the example of quantum gravity induced collapse model. It has been hypothesized by Ellis and co-workers that spatial superpositions should decay due to the interaction of the system (modeled by a matter field) with wormholes. On a formal level, the calculation carried out in Ref. \cite{Ellis_1989} uses the scattering matrix approach (i.e. the same approach as in \cite{Joos_1985}) leading to the scaling of the decoherence rate with the separation $\Gamma \propto d^2$ provided $k_{\rm QG} d \ll 1$ ($k_{\rm QG}$ is the wave vector associated to the wormholes). In this regime one can expand the localization function to second order
\beq
  \Phi(d) \approx \Phi(0)+\frac{1}{2}\partial^2_d \left. \Phi(d) \right|_{d=0} d^2,
  \label{eq:expansion Phi}
\eeq
where the linear term is absent since $\Phi$ is an even function \cite{Bassi_2013}. Note that the expansion (\ref{eq:expansion Phi}) is well justified as $1/k_{\rm QG}$ corresponds to long wavelengths (up to $10^4$ m \cite{Ellis_1989}). Moreover, we focus on the case where $N a_{\rm latt} \ll d(t),$ i.e. the atom number times the lattice spacing is typically negligible with respect to the state separation. Those two considerations allow one to rewrite the master equation \eqref{eq:rho dot} as
\beq
  \dot \rho =  \frac{\Gamma_{\rm QG}}{2} \,  \left(   S_z \rho\, S_z-\frac{1}{2}  S_z^2\rho- \frac{1}{2} \rho  \,S_z^2 \right),
\eeq
where $S_z = \sum_{i=1}^N \sigma_z^i$ is the collective spin, i.e. the decoherence due to quantum gravity corresponds to a \emph{collective} $z$-spin noise, and the decay rate reads \cite{Ellis_1989} (see also Appendix \ref{sec:Decoherence})
\beq
	\Gamma_{\rm QG}= \frac{(c m_0)^4}{(\hbar m_{\rm Pl})^3} m_{\rm at}^2 d(t)^2 \equiv \gamma_{\rm QG} d(t)^2,
    \label{eq:decay unconv micro}
\eeq
where $c$ is the vacuum speed of light, $m_0$ the nucleon mass, $m_{\rm Pl}$ the Planck mass and $m_{\rm at}$ is the mass of a single trapped atom. Most importantly, it scales quadratically with the state separation, i.e. can be enhanced by increasing $d(t)$. Considering the initial product state (\ref{eq:prod state}), one can evaluate the overlap as (see Appendix \ref{sec:Decoherence})
\beq
	O_\text{QG}(t)=  \int {\rm d}\Lambda \cos^{2N} \!\!\big(\frac{\Lambda}{2} \big) \frac{e^{-\frac{\Lambda^2}{2 \gamma(t)} }}{\sqrt{2 \pi \gamma(t)}},
	\label{eq:overlap unconv}
\eeq
where $\gamma(t) = 2\gamma_{QG} \int_0^t d(t')^2 \, dt'$ is the variance of the gaussian distribution.\\

\noindent \emph{Scaling of coherences with atom number --} It is interesting to compare how the two overlaps $O_{\rm sc}, O_{\rm QG}$ scale with respect to the number of trapped atoms $N$.  While $O_\text{sc}$  decreases exponentially with $N$ \eqref{eq:overlap conv}, the $N$ dependence is more complicated for quantum gravity \eqref{eq:overlap unconv}. However, in the asymptotic limit of large $N$ (see Appendix \ref{sec:Decoherence}), $O_\text{QG}\to \sqrt{ \frac{1}{2 N\gamma(t)}} \vartheta _3 (0,e^{-\frac{(2\pi) ^2}{2 \gamma (t)}}), $ i.e. exhibits a decay with $N^{1/2}$ ($\vartheta _3$ is the Jacobi theta function). Consequently, we conclude that for product states of the form (\ref{eq:prod state}), it is preferable to use small atom numbers. 

Note, that the situation is completely different if the initial state is the $N$ atom GHZ state $\ket{\rm GHZ}=\frac{1}{\sqrt{2}}(\ket{g}^{\otimes N} + \ket{s}^{\otimes N})$. In this case, the decay of the coherence term $\coh{g}{s}^{\otimes N}$ scales as  $e^{-N^2}$ for quantum gravity and $e^{-N}$ for the photon scattering (see Appendix \ref{sec:Decoherence}). Hence, in this particular case, it is easier to test unconventional decoherence with large atom numbers. Although desirable, the creation of GHZ states with large $N$, is known to be a difficult task in practice and for the remainder of the article we focus on the more realistic case of a single atom.\\

\noindent \emph{Feasibility study --} We now give a detailed analysis of an experiment using a single atom. First, the overlaps (\ref{eq:overlap conv}) and (\ref{eq:overlap unconv}) both take the simple form $O_{\rm label} = \frac{1}{2}({\rm exp}(-\Gamma_{\rm label} \tau)+1)$, where "label=sc,QG". Here
\beq
	\Gamma_{\rm QG} = \frac{\gamma_{\rm QG}}{\tau} \int_0^\tau {\rm d}t \-\ d(t)^2 \equiv \gamma_{\rm QG} d_{\rm eff}^2,
	\label{eq:Gamma_QG t}
\eeq
where $\tau$ is the time it takes to spatially separate and relocalize the two states $\ket{g},$ $\ket{s}.$ We have introduced an effective distance $d_{\rm eff}$ which allows us to write the expression for the unconventional decay in a simple form $\Gamma_{\rm QG} \propto d_{\rm eff}^2$ for any atomic motion $d(t)$. 

\begin{figure}[h!]
\vspace*{0cm}
\hspace*{-0.2cm}
 \includegraphics[width=9cm]{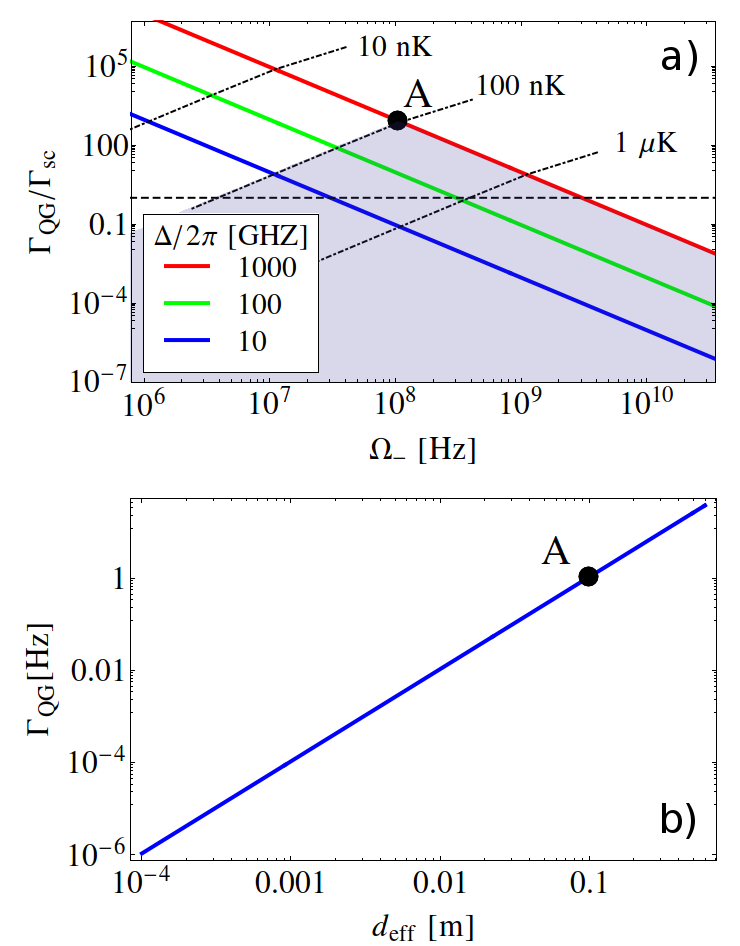}
   \vspace*{0cm}
\caption{(Color online) a) Ratio $r$ between the decoherence rates induced by quantum gravity and photon scattering as a function of the trapping Rabi frequency $\Omega_-$ for various trapping detunings $\Delta/(2 \pi)$ = \{10,100,1000\} GHz (blue, green, red lines respectively). The dashed black line is a guide for the eye corresponding to $r=1$. The dot-dashed lines represent different trap temperatures. Point A corresponds to the optimized ratio $r \approx 800$, see text for details. b) The decoherence rate due to quantum gravity (\ref{eq:decay unconv micro}) as a function of the distance $d_{\rm eff}$ for $^{87}$Rb. Point A corresponds to the distance $d_{\rm eff}=0.1$ m and $\Gamma_{\rm QG} = 1$ Hz respectively.} 
  \label{fig:single atom}	
 \end{figure}

It follows from (\ref{eq:Gamma_QG t}) that large spatial separations favor large $\Gamma_{\rm QG}.$ To get large separations by manipulating the $\ket{s}$ state independently, we consider a regime where $V^0_\pi \ll V^0_-$ which can be achieved already for moderate ratios between $\Omega_-$ and $\Omega_\pi$ (for example $\Omega_-/\Omega_\pi \approx 3$ yields $V^0_-/V^0_\pi \approx 10$). Furthermore, in order to have large separations in a short time, an acceleration ramp is required. Finding the exact atomic motion which maximizes the effect of the unconventional decoherence is a multidimensional optimization problem, as the motion depends on various parameters such as the Rabi frequency of the trapping laser, the detuning, the absolute timescale of the experiment, the temperature of the atoms or the maximal separation allowed by the experimental setup. To simplify the discussion, we consider a  smooth atomic motion described by 
\beq
	d(t)= d_{\rm max} + \frac{d_{\rm max}}{T} \left[ \frac{T}{2\pi}\sin{\frac{2\pi |t-T|}{T}} - |t-T| \right],
	\label{eq:distance fction}
\eeq
where $T \equiv \tau/2$ and $t \in [0,\tau]$, i.e. $d_{\rm max}$ is reached in half of the atomic round trip $\tau$ (see Appendix \ref{sec:Transport} for details). This leads to $d_{\rm eff} = d_{\rm max}/\sqrt{2}.$ To ensure that the atom follows the trapping potential, the trapping force needs to be larger than the dynamical force \cite{Schmid_2006}, $-\partial_x V_s > m_{\rm at} a$, where $a$ is the acceleration of the atom. Evaluating the trapping force at its maximum yields a constraint on the maximal acceleration for a given Rabi frequency and detuning
\beq	
	\frac{\hbar k}{m_{\rm at}} \frac{\Omega_-^2}{4 \Delta} \approx a_{\rm max} =  \frac{2\pi d_{\rm max}}{T^2},
	\label{eq:acc}
\eeq
where we have approximated $V_s(x,t) \approx V_-^0 \cos^2(kx+\varphi(t))$ in Eq. (\ref{eq:V_s}) and used the second derivative of Eq. (\ref{eq:distance fction}) to obtain the right hand side. 
Inspired by the experimental results presented in Ref. \cite{Schmid_2006}, where atomic transport was realized over up to 20 cm, we choose a more conservative value $d_{\rm eff} = 10$ cm. Fig. \ref{fig:single atom}a shows the ratio 
\beq
	r \equiv \Gamma_{\rm QG}/\Gamma_{\rm sc}
	\label{eq:ratio}
\eeq
as a function of $\Omega_-$ for various detunings $\Delta$. It can be seen that for large detunings and moderate Rabi frequencies $r \gg 1.$ In order to estimate the ratio $r$ that can be achieved in practice, we show in Appendix \ref{sec:Transport}, that $r$ scales as
\[
	r \propto d_{\rm max} k \Delta \tau^2, 
\]
i.e. one requires large $d_{\rm max}$, $\Delta$ and $\tau.$ 
First we note that $\Delta$ is bounded from above to make the coupling to the P$_{3/2}$ manifold negligible (as compared to the coupling to the $P_{1/2}$ manifold), i.e. to ensure a polarization selective atomic transport. We take $\Delta = 2 \pi \cdot 1$ THz. We then estimate the Rabi frequency $\Omega_-$ yielding the optimal $r$ from (\ref{eq:acc}). The optimal value of $\Omega_-$ depends on the duration of the experiment $\tau$. In order to get a quantitative idea, it is useful to estimate the timescale of the unconventional decoherence. $d_{\rm eff} = 10$ cm yields $\Gamma_{\rm QG} = 1$ Hz (point A in Fig. \ref{fig:single atom}b) and as an example, we choose $\tau = 1$ s. The optimal $\Omega_-$ maximizing $r$ is given by the point A in \fig{\ref{fig:single atom}a} ($\Omega_- \approx 10^8$ Hz, $r \approx 800$). Further increasing $\Omega_-$ (for all other parameters fixed) increases $\Gamma_{\rm sc}$ and reduces $r$. On the other hand, if $\Omega_-$ is decreased, the only possibility to reach $d_{\rm max}$ in a time $\tau/2$ is to decrease $\Delta$, which again increases $\Gamma_{\rm sc}$ and decreases $r$. The decrease of $r$ for deviations of $\Omega_-$ from the optimal value is represented by the shaded region in \fig{\ref{fig:single atom}}a.

To complete our discussion, specific values of $\Omega_-$ and $\Delta$ impose a maximal temperature of the atoms, such that they remain trapped \footnote{The equation (\ref{eq:Temperature}) refers to the trapping of the $\ket{s}$ state along the lattice axis. In practice the limiting temperature will be given also by the trapping force of the $\ket{g}$ state and in the radial direction.}. The dot-dashed lines in \fig{\ref{fig:single atom}}a show different trap temperatures, which are given by
\beq
  T_{\rm tr} = \frac{\hbar^\frac{3}{2}}{k_{\rm B}} \frac{\sqrt{2} k}{\sqrt{m_{\rm at}}} \frac{\Omega_-}{\sqrt{\Delta}},
  \label{eq:Temperature}
\eeq
where $k_{\rm B}$ is the Boltzmann constant. One can see, that the shaded region corresponds to $T_{\rm tr} \gtrsim 100$ nK, i.e. to the temperatures achievable in today's cold atomic experiments \cite{Dickerson_2013}. \\

\noindent \emph{Summary and Outlook --} 
To conclude, we have shown that unconventional decoherence models can be efficiently tested with cold atoms in superpositions of different spatial positions. When using optical lattices where the dominant source of decoherence is the photon scattering, the standard and quantum gravity based decoherence mechanisms operate differently: While the former acts as a local noise, the latter corresponds to a collective dephasing. Consequently, when dealing with product states, we have shown that it is easier to observe the unconventional decoherence with small atom numbers. We have performed a detailed feasibility study showing unambiguously that a single atom in a superposition of two positions separated by $\sim 10$ cm for $\sim 1$ s can be used to test quantum gravity induced collapse. This proposal might be implemented using various platforms where atomic transport has been successfully demonstrated, including hollow core fibers \cite{Renn_1995, Christensen_2008}, optical tweezers \cite{Beugnon_2007}, atomic chips \cite{Schumm_2005,WangYJ_2005,ShinY_2005,Hofferberth_2006} or free space optical lattices \cite{Schrader_2001, Kuhr_2003, Mandel_2003, Dotsenko_2005, Browaeys_2005, Schmid_2006}. As a first outlook, we emphasize that our results allow one to bound the model of Ellis and co-workers \cite{Ellis_1989} in any scenario where atoms are spatially delocalized over sufficiently large distances. It is interesting to quantify the bounds provided e.g. by atom interferometers with large momentum transfers, yielding large spatial separations \cite{McDonald_2013, Dickerson_2013}. In particular, we are currently evaluating \footnote{J. Min\'{a}\v{r}, P. Sekatski and N. Sangouard, in preparation} what are the contraints on the model of Ref. \cite{Ellis_1989} imposed by the experimental results published very recently in Ref. \cite{Kovachy_2015}. Another interesting perspective is to investigate the potential of many body entangled states, such as spin squeezed states \cite{Kuzmich_2000,Esteve_2008} or BEC solitons in guided interferometers \cite{Weiss_2009}, to benefit from a favourable scaling of the ratio between standard and unconventional decoherence rates, which we leave for future work. \\
~\\
\noindent \emph{Acknowledgments --} J.M. would like to thank T. Lahaye, L. Hackerm\"{u}ller, P. Kr\"{u}ger and I. Lesanovsky for useful discussions. N.S. thanks P. Treutlein and D. Meschede for stimulating discussions. J.M. was supported by the grant EU-FET HAIRS 612862. P.S. and N.S were supported by the Swiss National Science Foundation grant number P2GEP2\_151964 and PP00P2\_150579 respectively.




\appendix

\begin{widetext}

\section{Analysis of Decoherence mechanisms}
\label{sec:Decoherence}

\subsection{Conventional decoherence due to photon scattering}

\paragraph{Master equation.} Here we introduce the master equation governing the decoherence due to the scattering of the optical trapping photons. It is a well known fact, that in the far off resonant regime where  operate optical lattices, the induced excited state population can be typically neglected (in fact is proportional to $(\Omega/\Delta)^2$, where $\Omega$ is the Rabi frequency of the trapping laser and $\Delta$ its detuning from the excited state, which is easy to show e.g. from the Bloch equations). This is a desired feature, since one typically wants to avoid the heating due to the photon absorption/emission by the trapped atoms. As a consequence, one can consider only the decay of coherences (which are of order $\Omega/\Delta$) while neglecting the change in the state population. Next we consider a situation, where the individual sites of the optical lattice can be spatially fully resolved by measuring the scattered photons. This is indeed the feature which led to the success of and is exploited in the quantum gas microscopes \cite{Bakr_2009,Sherson_2010}. Under these assumptions, the decoherence of the system reduces to the product of individual decoherence events, where a measurement of a scattered photon results in a projection of the atom to a given spin state of the ground state manifold. The corresponding master equation reads
\eq
\dot\rho =  2 \sum_{i=1}^N \Big( \Gamma_g\big( \prjct{g}_i \rho \prjct{g}_i- \frac{1}{2} \{\prjct{g}_i,\rho\}\big) + \Gamma_s\big( \prjct{s}_i \rho \prjct{s}_i- \frac{1}{2} \{\prjct{s}_i,\rho\}\big)\Big),
\eeq
where $\Gamma_\alpha$, $\alpha=g,s$, are the scattering rates (\ref{eq:decay conv}) for the atoms in $\ket{g}$, $\ket{s}$ state and $\{\cdot,\cdot\}$ is the anticommutator. Straigtforward algebra allows to rewrite this expression  in terms of the operators $\sigma_z^i =\prjct{g}_i-\prjct{s}_i$ yielding 
\beq\label{ME conv}
\dot \rho = \sum_{i=1}^N \frac{\Gamma_g+\Gamma_s}{2}\big( \sigma_z^i \rho \, \sigma_z^i-\rho \big)
\eeq

~\\
\paragraph{State evolution.} It is straightforward to solve the master equation \eqref{ME conv}. The dephasing process acts locally resulting in the decay of the off-diagonal terms. Formally this process can be described by the completely positive trace preserving (CPTP) map $\mathcal{E}^{\rm loc}_t(\rho(0)) = \rho(t)$, where $\mathcal{E}^{\rm loc}_t=\bigotimes_{i=1}^N \varepsilon_t^i$ is given simply by the product of the $N$ individual dephasing processes
\beq\label{CP conv}
\varepsilon_t^i(\rho) = \left(\frac{1+e^{-(\Gamma_g+\Gamma_s)t}}{2}\right)\rho +\left(\frac{1-e^{-(\Gamma_g+\Gamma_s)t}}{2}\right) \sigma_z^i \rho \sigma_z^i.
\eeq
For the initial product state (\ref{eq:prod state}), $\ket{\psi_0}=\ket{+}^{\otimes N}=\left(\frac{\ket{g}+\ket{s}}{\sqrt{2}}\right)^{\otimes N}$, the overlap with a state at a later time is given by
\beq
O_\text{sc}= {\rm Tr}\left(\prjct{\psi_0} \rho(t)\right)= \prod_{i=1}^N \bra{+} \varepsilon^i_t\left( \ket{+}\bra{+} \right) \ket{+}=\left(\frac{1+e^{-(\Gamma_g+\Gamma_s)t}}{2}\right)^N
\eeq

\subsection{Unconventional decoherence due to quantum gravity}

\paragraph{Master equation.} 
As described in the main text, the spatial decoherence can be described phenomenologically by the master equation \cite{Bassi_2013}
\beq
	\dot \rho  =  - \int d{\bf x'} d{\bf x} \,\Gamma({\bf x}, {\bf x'}) \coh{\bf x}{\bf x} \rho \coh{\bf x'}{\bf x'}
	\label{eq:rho dot app}
\eeq
where $\ket{{\bf x}}=\ket{x_1,...,x_N}$  is the position basis for the $N$ particles. We start by explicitly computing the localization rate (\ref{eq:Gamma})
\beq
\Gamma({\bf x}, {\bf x'}) = \frac{\gamma_0}{2} \sum_{i,j=1}^N \mu_i \mu_j  \Big( \Phi(x_i - x_j)+\Phi(x_i' - x_j')- 2\Phi(x_i - x_j')\Big).
\eeq
The localization function $\Phi$ can be expanded for small separations $d$ compared to the localization length $L$, which is well justified when considering the model of \cite{Ellis_1989}, where $L \sim 10^4$ m and we consider separations of $d \sim$ 10 cm. With this assumption $\Phi(d) \approx \Phi(0) + \frac{1}{2} \Phi''_0 d^2$, where $\Phi''_0 = \partial^2_d \left. \Phi(d) \right|_{d=0}$ and we have used the fact, that $\Phi$ has to be symmetric, $\Phi(d) = \Phi(-d)$, which imposes the cancellation of the linear term in the expansion. 

Next, note that in the situation where the atoms are tightly trapped in the lattice, the spin state of each atom $\ket{\ell}_i$ is fully entangled with its position $x_i(\ell)$, such that the decoherence rate only depends on the global atomic state through $\Gamma({\bf l}, {\bf l}')$, where ${\bf l}=(\ell_1, ...,\ell_N)$. Denoting the spin states $\ket{0} \equiv \ket{g}$ and $\ket{1} \equiv \ket{s}$, for lattice spacing $a_{\rm latt}$ and separation $d(t)$ the position of each atom is given by $x_i(\ell) = i a_{\rm latt}  + \ell d(t)$. In the regime of interest of large separation $Na_{\rm latt} \ll d(t)$, the separation of two atoms within the same lattice can be neglected and $x_i(\ell) \approx \ell d(t)$. In this approximation and putting $\mu_i = \mu$ for all $i$ as we consider identical particles
\begin{align}
\Gamma({\bf l}, {\bf l}') =  & \frac{\gamma_0 \,\mu^2  \, \Phi''_0 d(t)^2}{4} \sum_{i,j=1}^N  \Big( (\ell_i-\ell_j)^2 + (\ell_i'-\ell_j')^2-2(\ell_i-\ell_j')^2\Big)\nonumber \\
= & \frac{\gamma_0 \,\mu^2  \, \Phi''_0 d(t)^2}{4} \Big(4 (\sum_i \ell_i)(\sum_i \ell_i')-2(\sum_i \ell_i)^2-2(\sum_i \ell_i')^2\Big) \nonumber\\
= & \frac{\gamma_0 \,\mu^2  \, \Phi''_0 d(t)^2}{4} \Big(S_z({\bf l})S_z({\bf l}')-\frac{1}{2}S_z({\bf l})^2-\frac{1}{2} S_z({\bf l}')^2 \Big)
\end{align}
where $S_z({\bf l})=\sum_i (2 \ell_i -1)$ is the total z-spin of the state $\ket{l}$, $S_z \ket{\bf l}= S_z({\bf l})\ket{\bf l}$, with $S_z=\sum_i \sigma_z^i= \sum_{\bf l}S_z({\bf l})\prjct{{\bf l}}$. This allows to rewrite the master equation (\ref{eq:rho dot app}) governing the localization in terms of total spin
\beq\label{ME unconv}
\dot \rho = \frac{\gamma_0 \,c^2  \, \Phi''_0 d(t)^2}{4} \Big(S_z \rho S_z - \frac{1}{2}\{S_z^2,\rho\}\Big).
\eeq
The prefactor in (\ref{ME unconv}) can be linked to the parameters of \cite{Ellis_1989} by combining the results of \cite{Ellis_1989} with the present analysis. This was done e.g. in \cite{RomeroIsart_2011} with the result
$ \frac{\gamma_0 \,\mu^2  \, \Phi''_0 d(t)^2}{2} =\Gamma_\text{QG}= \frac{(c m_0)^4}{(\hbar m_{\rm Pl})^3} m_{\rm at}^2 d(t)^2$.

~\\
\paragraph{State evolution.} In order to solve the master equation\eqref{ME unconv} we realize that the collective spin noise process corresponds to the diffusion of the polar angle of the total spin. After an elementary time step, the density matrix can be written as an average over a random variable $\lambda$
\beq
\rho(t+dt)= \rho(t) + dt\,\frac{\Gamma_\text{QG}}{2} \left( S_z \rho(t) S_z - \frac{1}{2} S_z^2 \rho(t) - \frac{1}{2} \rho(t) S_z^2 \right) = \mn{e^{i \lambda S_z} \rho_t e^{-i \lambda S_z}}_{\lambda}
\label{eq:rho time step}
\eeq
where $\mn{\cdot}$ denotes an ensemble average over $\lambda$ with the moments $\mn{\lambda}=0$, $\mn{\lambda^2}=\Gamma_\text{QG} \,dt/2$ and $\mn{\lambda^{(n>2)} }=\mathcal O(dt^2)$. This can be easily verified by substituting the expansion of the operator ${\rm exp}(i \lambda S_z) = 1 + i \lambda S_z -\frac{1}{2}\lambda^2 S_z^2 + \mathcal{O}(\lambda^3)$ into (\ref{eq:rho time step}). After $n=\frac{t}{dt}$ time steps the state of the system is 
\beq
\rho(t) = \mathcal{E}^{\rm col}_{t}(\rho(0))=\mn{e^{i (\sum_{i=1}^n \lambda_i ) S_z} \rho(t) e^{-i ( \sum_{i=1}^n \lambda_i ) S_z}}_{\lambda_1,...,\lambda_n} = \int {\rm d}\Lambda e^{i \Lambda S_z} \rho(0) e^{-i \Lambda S_z} \frac{e^{-\Lambda^2/( 2\gamma(t) )}}{\sqrt{2\pi \gamma(t)}},
\eeq
where we have introduced the CPTP map $\mathcal{E}^{\rm col}_t$ corresponding to the collective noise. 

In the last equality we have used the fact that for $dt\to 0$ the sum of independent random variables $\Lambda= \sum_{i=1}^n \lambda_i$ is a normally distributed random variable with the variance $\gamma(t) = 2\gamma_{QG} \int_0^t d(t')^2 dt'$.

The overlap of the state at time $t$ with the initial state (\ref{eq:prod state}) can now be easily computed as follows. First we write the expression for the overlap
\beq
O_\text{QG} = {\rm Tr}\left( \prjct{\psi_0} \rho(t) \right)= \int {\rm d}\Lambda |\bra{\psi_0}e^{i \Lambda \sigma_z}\ket{\psi_0}|^{2N} \frac{e^{-\Lambda^2/( 2\gamma(t) )}}{\sqrt{2\pi \gamma(t)}} = 
\int {\rm d}\Lambda \cos^{2N}\big(\frac{\Lambda}{2}\big) \frac{e^{-\Lambda^2/( 2\gamma(t) )}}{\sqrt{2\pi \gamma(t)}}.
\label{eq:O QG expr}
\eeq
The evaluation of the integral can be performed numerically. One can proceed further in the large $N$ limit - to get the assymptotic expansion for $O_\text{QG}$ we remark that
\beq
\cos^{2N}(\frac{\Lambda}{2})\to \sqrt{\frac{\pi}{N}}\sum_{k=-\infty}^\infty \!\!\delta(\Lambda+ 2 \pi k)  \quad \text{for} \quad N\to \infty,
\label{eq:expr}
\eeq
where $\delta$ is the Kronecker delta and we have used 
\beq
	\int_{-\frac{\pi }{2}}^{\frac{\pi }{2}} \text{cos}^{2 N}\left(\frac{\Lambda }{2}\right) \, d\Lambda =\frac{\pi  \binom{2 N}{N}}{2^{2 N}}\to \sqrt{\frac{\pi }{N}}.
\eeq
Substituting (\ref{eq:expr}) back to (\ref{eq:O QG expr}) yields an infinite sum, which can be evaluated as $\sum_{k=-\infty}^\infty e^{-\frac{(2\pi k)^2}{2\gamma}} = \vartheta _3\left(0,e^{-\frac{(2 \pi )^2}{2\gamma }}\right)$, where $\vartheta_3$ is the Jacobi $\vartheta$ function. This leads to the result
\begin{align}
O_\text{QG} \to \sqrt{\frac{1}{2 N \gamma(t) }}\, \vartheta _3\left(0,e^{-\frac{(2 \pi )^2}{2\gamma(t) }}\right) \quad \text{for} \quad N\to \infty.
\end{align}

\subsection{GHZ states}

It is interesting to compare the decay of the overlap of the product state (which we considered for the simplicity of its experimental preparation) with the decay for some highly entangled state. A benchmark example in metrology is the GHZ state $\rho_{\rm GHZ} = \ket{\rm GHZ} \bra{\rm GHZ}$, $\ket{\rm GHZ}=1/\sqrt{2}(\ket{g}^{\otimes N}+\ket{s}^{\otimes N})$. One gets for the local and collective dephasing respectively
\begin{align}
\mathcal{E}_t^{\rm loc}\big( \rho_{\rm GHZ} \big)&=\frac{1}{2}\Big( \prjct{g}^{\otimes N}+\prjct{s}^{\otimes N} +\big(\coh{g}{s}^{\otimes N} +{\rm h.c.} \big) e^{-N(\Gamma_g+\Gamma_s)t}  \Big) \\
\mathcal{E}_t^{\rm col}\big( \rho_{\rm GHZ} \big)&=\frac{1}{2}\Big( \prjct{g}^{\otimes N}+\prjct{s}^{\otimes N} +\big(\coh{g}{s}^{\otimes N} +{\rm h.c.} \big)e^{-2 N^2\gamma(t)}  \Big),
\end{align}
which yields the overlaps
\begin{align}
O_\text{sc}^{\rm GHZ}&= \bra{\rm GHZ}\mathcal{E}_t^{\rm loc}\big( \rho_{\rm GHZ} \big) \ket{\rm GHZ}= \frac{1}{2}(1 + e^{-N(\Gamma_g+\Gamma_s)t} ) \\
O_\text{QG}^{\rm GHZ}&= \bra{\rm GHZ}\mathcal{E}_t^{\rm col}\big( \rho_{\rm GHZ} \big) \ket{\rm GHZ}= \frac{1}{2}(1 + e^{-2 N^2\gamma(t)} ).
\end{align}

\section{Atomic transport}
\label{sec:Transport}

In this section we provide a simple description of the atomic motion in a time dependent optical lattice. Considering the tight binding situation, where the atoms are firmly trapped at an intensity extrema of the lattice, it is in principle possible to displace the atoms by an arbitrary distance $d$. Working in the regime $V^0_- \gg V^0_\pi$, the $s$ species can be displaced independently of the $g$ species and thus, in principle, allows for creation of a superposition of type (for a single atom)
\beq
	\ket{\psi} = c_g \ket{g} \psi(x-x_g) + c_s \ket{s} \psi(x-x_s), 
\eeq
where we assume, that due to the tight binding the position of the atom is treated classically and can be described by a classical field $\psi(x)$, which has the property $\int {\rm d}x \, x |\psi(x-x_\alpha)|^2 = \int {\rm d}x x \delta(x-x_\alpha)=x_\alpha$, where $\alpha=g,s$. In other words, with respect to the position in real space we treat the atom as a classical point like object, which is a justified description for the system where separations of multiples of lattice spacings between $\ket{g}$ and $\ket{s}$ are achieved.

As only the $s$ species is displaced, we thus seek solutions of the classical equation of motion associated with the Hamiltonian
\beq
	H = \frac{p_s^2}{2m_{\rm at}} + V_s(x,t),
	\label{eq H}
\eeq
where $V_s(x,t) \approx V_-^0 \cos^2(kx+\varphi(t))$ is given by (\ref{eq:V_s}). The Hamilton equations of motion given by (\ref{eq H}), $\dot{p} = \partial_x H$, $\dot{x} = -\partial_p H$, can be combined to give
\beq
	\ddot{x}_s = -\frac{V^0_- k}{m_{\rm at}} \sin(2k x_s + 2 k x_{\rm latt}(t)),
	\label{eq:EoM}
\eeq
where $x_{\rm latt}(t)$ is the lattice motion. As explained in the main text, the specific choice of the lattice motion is a result of an optimization with respect to a given figure of merit and subject to the constraints such as a maximal achievable separation or the timescale of the experiment. For the sake of concreteness and motivated by the experimental results of \cite{Schmid_2006}, we consider lattice motion
with an initial sinusoidal acceleration ramp such that the maximal separation $d_{\rm max}$ is reached after a time $T$. When $d_{\rm max}$ is reached, the atoms remain separated for a waiting time $t_w$ before they are brought back together with the same acceleration ramp of duration $T$,
\beq
  x_{\rm latt}(t)=\begin{cases} 
      \frac{d_{\rm max}}{T} \left[ \frac{T}{2\pi}\sin{\frac{2\pi (T-t)}{T}} - (T-t) \right] & t < T \\
      d_{\rm max} & T \leq t \leq t_w+T \\
d_{\rm max}-\frac{d_{\rm max}}{T} \left[ \frac{T}{2\pi}\sin{\frac{2\pi (t-T-t_w)}{T}} - (t-T-t_w) \right] & t_w+T < t \leq t_w+2T.
   \end{cases}
   \label{eq:motion 1}
\eeq

In order to control the atomic motion, one requires the atoms remain trapped  in a given lattice site at all times, i.e. that the trapping force is larger than the dynamical force $m_{\rm at} \ddot{x}_s$ acting on the atoms. It is easy to check, that the relation (\ref{eq:acc}) provides indeed the desired condition. By numerically solving (\ref{eq:EoM}) we have verified, that when the acceleration of the lattice $a \lesssim a_{\rm max}$, the atoms follow the lattice potential, while they are not able to follow for $a > a_{\rm max}$. In the following, we thus restrict only to the situation, where $a \lesssim a_{\rm max}$ and the atomic motion $d(t)$ can be identified with the lattice motion (\ref{eq:motion 1}).

\subsection{Optimization of the atomic motion}

Next, we would like to optimize the lattice motion of the form (\ref{eq:motion 1}), i.e. optimizing $T$ and $t_w$ with respect to the ratio $r=\Gamma_{\rm QG}/\Gamma_{\rm sc}$ between the unconventional and conventional decoherence rates, eq. (\ref{eq:ratio}). Since $\Gamma_{\rm QG} \propto \int_0^\tau {\rm d}t d(t)^2$, where $\tau$ is the total duration of the atomic round trip, one can define an effective distance $d_{\rm eff}$ for any atomic motion $d(t)$ and $\tau$ through
\beq
	d_{\rm eff}^2 \tau \equiv \int_0^\tau {\rm d}t \, d(t)^2.
	\label{eq:d eff}
\eeq
Taking $d(t) = x_{\rm latt}(t)$, eq. (\ref{eq:motion 1}), one readily finds
\beq
	d_{\rm eff}^2 = d_{\rm max}^2 \left[ 1 + \frac{2 T}{\tau} (\alpha-1)\right],
\eeq
where $\tau = 2T + t_w$ and
\beq
	\alpha = \frac{1}{d_{\rm max}^2 T} \int_0^T {\rm d}t \, d(t)^2 = \frac{1}{24}\left( 8 + \frac{15}{\pi^2}\right) \approx 0.4.
\eeq
Since $\Gamma_{\rm QG} \propto d_{\rm eff}^2$ and $\Gamma_{\rm sc} \propto \Omega^2/\Delta^2$,
\beq
	r \propto \frac{d_{\rm eff}^2 \Delta^2}{\Omega^2}.
	\label{eq:ratio prop}
\eeq
The maximal possible acceleration for a given Rabi frequency is given by (\ref{eq:acc}) and at the same time is related to the motion (\ref{eq:motion 1}) by
\[
	a_{\rm max} = \frac{2\pi d_{\rm max}}{T^2}.
	\label{eq:acc T}
\]
Substituting (\ref{eq:acc}) to (\ref{eq:ratio prop}) yields
\beq
	r \propto d_{\rm max} k \Delta \left( 1 + 2(\alpha-1)\frac{T}{\tau}\right) T^2, 
\eeq
where we have kept all the factors depending on the separation, detuning or time. The ratio $r$ is thus maximized for maximum possible detuning $\Delta$, separation $d_{\rm max}$ and, subject to the constraint $T \leq \tau/2$, for $T=\tau/2$. Using $T=\tau/2$ (i.e. $t_w=0$) in (\ref{eq:motion 1}) then yields the optimized atomic motion (\ref{eq:distance fction}).



\end{widetext}

\end{document}